\documentstyle[twocolumn,epsf]{jpsj}

\def\runtitle{Non-Fermi liquid behavior in CeNi$_2$Ge$_2$}
\def\runauthor{Yuji~{\sc Aoki}}

\title{Specific Heat Study of Non-Fermi Liquid Behavior in CeNi$_2$Ge$_2$: Anomalous Peak in Quasi-Particle Density-of-States }

\author{Yuji~{\sc Aoki}\footnote{E-mail: aoki@phys.metro-u.ac.jp}, Jun~{\sc Urakawa}, Hitoshi~{\sc Sugawara}, Hideyuki~{\sc Sato}, Tadashi~{\sc Fukuhara}$^1$ and Kunihiko~{\sc Maezawa}$^1$}

\inst{Department of Physics, Faculty of Science, Tokyo Metropolitan University, Minami-Ohsawa 1-1, Hachioji, Tokyo 192-03\\
$^1$Faculty of Engineering, Toyama Prefectural University, Kosugi, Toyama 939-03\\}

\recdate{July 8, 1997} % ~~~~~~~~~~~~~~~~~~~~~

\abst
{
To investigate the non-Fermi liquid (NFL) behavior in a nonalloyed system CeNi$_2$Ge$_2$, we have measured the temperature and field dependences of the specific heat $C$ on a CeNi$_2$Ge$_2$ single crystal. 
The distinctive temperature dependence of $C/T$ ($\sim\alpha-\beta\sqrt{T}$) is destroyed in almost the same manner for both field directions of $B$//c-axis and $B$//a-axis.
The overall behavior of $C(T,B)$ and the low-temperature upturn in $\chi(T)$ can be reproduced, assuming an anomalous peak of the quasi-particle-band density-of-states (DOS) at the Fermi energy possessing $\sqrt{\epsilon}$ energy dependence.
Absence of residual entropy around $T=0$ K in $B \sim 0$ T has been confirmed by the magnetocaloric effect measurements, which are consistent with the present model.
The present model can also be applied to the NFL behavior in CeCu$_{5.9}$Au$_{0.1}$ using a ln($\epsilon$)-dependent peak in the DOS.
Possible origins of the peak in the DOS are discussed.
}

\kword{CeNi$_2$Ge$_2$, non-Fermi liquid, heavy fermion, spin fluctuation, specific heat, magnetocaloric effect, magnetic susceptibility}

\begin{document}
\sloppy
\maketitle
\newpage

% \section{Introduction}
% NFL
In recent years, deviations from expected Fermi-liquid behavior observed in several {\it f}-electron Kondo-lattice compounds at low temperatures are attracting a great deal of attention in the area of strongly correlated electron systems. 
In the intensively studied systems, such as CeCu$_{6-x}$Au$_x$,~\cite{rf:Lo94CeCu6,rf:Bo95CeCu6} Ce$_{1-x}$La$_x$Ru$_2$Si$_2$,~\cite{rf:Ka96SCR} CeCu$_2$Si$_2$~\cite{rf:St96NFL} and Ce$_7$Ni$_3$,~\cite{rf:Um96CN} physical quantities, such as specific heat $C$, electrical resistivity $\rho$ and magnetic susceptibility $\chi$, show anomalous temperature dependences on the verge of long-range magnetic ordering, when the relevant parameters are tuned by changing atomic concentration or pressure.
The commonly observed behaviors are $C/T\propto$-ln($T/T_0$) or $a-b\sqrt{T}$, $\Delta\rho \propto T^n$ with $n<2$ and $\chi=a'-b'\sqrt{T}$.
At the moment, as possible origins of such non-Fermi-liquid (NFL) behaviors, (1) a strong quantum effect of the low-energy spin fluctuation (SF) near a $T=0$ magnetic phase transition,~\cite{rf:Mo95SCR,rf:Mi93NFL} and (2) a distribution of Kondo temperature $T_K$,~\cite{rf:Mi97TKdist} have been proposed.
For studying the former effect, randomness inevitably introduced in the alloyed systems and complexity due to more-than-one crystallographic 4{\it f}-site in the unit cell should be avoided.
Therefore, a search for NFL behaviors in stoichiometric systems with only one crystallographic 4{\it f}-site is highly desirable.
Examples of desirable systems reported to date are CeNi$_2$Ge$_2$,~\cite{rf:St96NFL} CeCu$_2$Si$_2$~\cite{rf:St96NFL} and CePd$_2$Si$_2$,~\cite{rf:Gr96CPS} all of which have tetragonal ThCr$_2$Si$_2$-type crystal structure.
% CeNi2Ge2
CeNi$_2$Ge$_2$ was first reported as a paramagnetic heavy-fermion compound with the electronic specific heat coefficient of $\gamma \sim350$ mJ/K$^2$mol.~\cite{rf:Kn88CNG}
Recently, the gradually increasing $C/T$ with decreasing temperature and a deviation from $\rho \propto T^2$ behavior at ambient pressure have attracted attention as NFL behaviors.~\cite{rf:St96NFL}
In the alloying experiment of Ce(Ni$_{1-x}$Cu$_x$)$_2$Ge$_2$,~\cite{rf:Lo92Cudope,rf:Kn97Cudope} the end compound CeNi$_2$Ge$_2$ demonstrated that it is close to the critical concentration of $x_c\sim0.2$, above which the antiferromagnetic ordering sets in.
Therefore, in CeNi$_2$Ge$_2$, located around the boundary between the paramagnetic heavy-fermion and magnetically ordered regimes, antiferromagnetic spin fluctuation is expected to play a role in NFL behaviors.
A metamagneticlike anomaly observed at 42 T in the magnetization curve and a broad maximum in $\chi$ for $B//$c-axis are regarded as evidence of antiferromagnetic correlation.~\cite{rf:Fu96Meta} 
In this letter, we report precise measurements of specific heat as functions of both temperature and magnetic field on a single crystal CeNi$_2$Ge$_2$ to investigate the low-energy excitations which should be related to the NFL behaviors.

% \section{Experimental}
The preparation of a single-crystalline sample was described elsewhere.~\cite{rf:Fu96Meta}
The specific heat was measured by a semiadiabatic heat-pulse method using a dilution refrigerator in the magnetic fields up to 8 T.
In the measurement, the heat capacity of the addenda, which has been carefully subtracted from the measured total heat-capacity, amounts to $2\%$ (19$\%$) of the sample heat-capacity at 0.2 K (7 K) in zero field. 

% \section{Results and Discussion}
The temperature dependence of the specific heat measured in the fields of $B//$c-axis and $B//$a-axis is shown in Fig.\ref{fig:CBTT} plotted as $C/T$ vs $\sqrt{T}$.
The lattice contribution generally expressed as $C_L = \delta T^3$ should be negligibly small in the measured temperature range; $\delta=0.295$ mJ/K$^4$mol estimated from LaRu$_2$Si$_2$~\cite{rf:Be85Cp} leads to $C_L/T\sim0.01$ J/K$^2$mol at 6 K, which amounts to 4\% of the measured value in zero field.
% Šj"ä"M
The nuclear magnetic specific heat, estimated to be 6$\times 10^{-8} (B/T)^2$ J/Kmol in the field of $B$, is also negligible.
Thus, the observed $C$ is attributed mainly to the electronic contribution.
As a whole, no essential difference has been found between the field directions.
The data in zero field can be well fitted by $C/T=\alpha-\beta\sqrt{T}$ with $\alpha=$0.410 J/K$^2$mol and $\beta=$0.077 J/K$^{2.5}$mol. % (2)(1)
This behavior is similar to that in the data for a polycrystalline sample reported by Steglich {\it et al.},~\cite{rf:St96NFL} although their high-$T$ data are expressed well by $C/T \propto$ ln$(T/T_0)$.
With increasing field, $C/T$ decreases gradually below $\sim$1 K.
In the fields of $B>$4 T, almost $T$-independent $C/T$ appears at low temperatures and $C/T$ shows a broad maximum at $T_{max}$, which is plotted in a $B$ vs $T$ phase diagram in Fig. \ref{fig:BT}.
Note that $T_{max}$ is close to the temperature $T_l$ where the resistivity starts to deviate from the low-temperature $T^2$ dependence.~\cite{rf:St96NFL}
Contrary to low temperatures, $C/T$ increases with increasing field at high temperatures.
As shown in Fig.~\ref{fig:CBTB}, $C/T$ as a function of field changes sensitively around zero field at low temperatures.
It should be noted that the field dependence of $C/T$ for $T\to0$ can be expressed by $C/T=0.410-0.046\sqrt{B}$ J/K$^2$mol. % (2) (1)

%%%   model
In this letter, we demonstrate that both $\sqrt{T}$-dependence of $C/T$ in zero field and the $\sqrt{B}$-dependence of $C/T$ for $T\to0$ can be simply understood by assuming an anomalous peak structure in the quasi-particle density-of-states (DOS) as a function of energy.
In general, $C/T$ can be expressed as
\begin{equation}
\label{eq:CBT}C/T=2\int_{-\infty}^{\infty} dx \bigl[D_{\uparrow}(\epsilon_{+})+D_{\downarrow}(\epsilon_{-}) \bigr] x^2{\rm sech}^2(x) ,
\end{equation}
where $\epsilon_{\pm}=2Tx\pm(\mu_{eff}/k_B)B$, and $D_{\uparrow}(\epsilon)$ and $D_{\downarrow}(\epsilon)$ (in the unit of states$\cdot$J/K$^2$mol$\cdot$spin) represent the DOS for up-spin and down-spin bands, respectively.
Here we assume rigid quasi-particle-bands without temperature dependence expressed as
\begin{equation}
\label{eq:DOS}D_{\uparrow}(\epsilon)=D_{\downarrow}(\epsilon)=\alpha '-\beta '\sqrt{|\epsilon|}.
\end{equation}
Combined with eq.~(\ref{eq:CBT}), this model leads to the temperature dependence of $C/T$ in $B=0$ and the field dependence of $C/T$ at $T=0$ as follows,
\begin{equation}
\label{eq:CBTB0}C/T=(\frac{2\pi^2}{3})\alpha '-\bigl[\frac{15}{8}(4-\sqrt{2})\sqrt{\pi}\zeta(\frac{5}{3}) \bigr]\beta '\sqrt{T} \ \ \ \ ({\rm for} \, B=0)
\end{equation}
\begin{equation}
\label{eq:CBTT0}C/T=(\frac{2\pi^2}{3})\bigl[\alpha '-\beta '\sqrt{|\frac{\mu_{eff}}{k_B}B|}\bigr] \ \ \ \ ({\rm for} \, T=0), 
\end{equation}
where $\zeta(x)$ is the Zeta function.
Comparing eq.~(\ref{eq:CBTB0}) with the experimental results, the parameters in eq.~(\ref{eq:DOS}) are determined to be $\alpha '=6.23\times10^{-2}$ states$\cdot$J/K$^2$mol$\cdot$spin and $\beta '=6.68\times10^{-3}$ states$\cdot$J/K$^{2.5}$mol$\cdot$spin.
Then, inserting the $C/T$ value at $B=8$ T into eq.~(\ref{eq:CBTT0}), we can estimate the effective magnetic moment $\mu_{eff}$=1.63 $\mu_B$.
Using the set of obtained parameters $\alpha ', \beta '$ and $\mu_{eff}$, we calculated $C/T$ as functions of both temperature (Fig.\ \ref{fig:CBTTcalc}) and field (Fig.\ \ref{fig:CBTBcalc}).
The broad maxima in $C/T$ vs $T$ above 4 T are well reproduced, and the calculated $T_{max}(B)$ is close to the experimental ones, as shown in Fig.\ \ref{fig:BT} with a broken line.
In the model, this broad peak originates from the thermal excitations to the peaks in the DOS shifted from the Fermi energy by $\mu_{eff}B$.
Above $\sim$3 K, the calculated $C/T$ gradually increases with increasing field, which is also consistent with the experimental results.
Quantitatively, however, there remain some discrepancies, i.e., the calculated $C/T$ for $B=$1 and 2 T becomes constant as $T\to0$, whereas the experimental $C/T$ increases continuously with decreasing $T$.
Although $T_{max}$ is proportional to $B$ in the calculation, the observed $T_{max}$ in Fig.\ \ref{fig:BT} shows subtle upward curvature, indicating a gradual change of $\mu_{eff}$. 
These discrepancies suggest that the peak structure of the quasi-particle DOS is not ''rigid'' but depends on $T$ and/or $B$ to some extent, especially close to the top of the peak.

%%% Entropy S
In many Ce compounds exhibiting the NFL ground state, the entropy related to the NFL behavior is reduced to nearly $1/2$ of the $R$ln2 expected for a doublet ground state.~\cite{rf:Se97CeNFL}
This fact leads to a naive but interesting question, i.e., whether any residual entropy is left at approximately $T=0$ in the NFL ground state or not.
If so, the entropy which is concealed below 0.2 K could be released at higher temperatures in the fields above 4 T, where Fermi liquid behaviors are recovered. 
In order to reveal any concealed entropy, we measured the magnetocaloric effect (MCE).
In the data analysis, we use the thermodynamical relation
\begin{equation}
\label{eq:MCE}\bigl( \frac{\partial S}{\partial B} \bigr)_T \simeq - \bigl( \frac{C}{T} \bigr) \bigl( \frac{\Delta T}{\Delta B} \bigr) .
\end{equation}
Here, $\Delta T/\Delta B$ is defined as $\Delta T/\Delta B= \frac{1}{2} (\Delta T_\uparrow - \Delta T_\downarrow) / | \Delta B | $ using the temperature change $\Delta T_\uparrow$ ($\Delta T_\downarrow$) after a slow increase (decrease) in the magnetic field by $|\Delta B|=0.02 \sim 0.1$ T.
This procedure eliminates the contribution from the eddy-current heating of the sample.
By integrating eq.~(\ref{eq:MCE}), i.e., $\int ( \frac{\partial S}{\partial B} )_T dB$, we obtained the field dependence of entropy at 0.5 K.
As shown in Fig.\ \ref{fig:SB}, the results are in good agreement with those calculated from $C(T)$ data, confirming that there is no concealed anomalous entropy below 0.2 K in the fields of $B\sim0$ T.

% magnetic susceptibility
The present model can be checked using the magnetic susceptibility data $\chi \equiv M/B$.
As shown in Fig.~\ref{fig:kaiT}, an upturn is seen below 10 K in both field directions,~\cite{rf:Fu96Meta} and was at first suspected to be a magnetic impurity contribution.
Although Steglich {\it et al.} pointed out the possible relevance of the upturn to the NFL behavior in $C$,~\cite{rf:St96NFL} conclusive evidence has not yet been provided.
Using the general formula
\begin{equation}
\label{eq:M}M=\frac{\mu_{eff}}{R}\int_{-\infty}^{\infty} \bigl[D_{\uparrow}(\epsilon+\frac{\mu_{eff}}{k_B}B)-D_{\downarrow}(\epsilon-\frac{\mu_{eff}}{k_B}B) \bigr] \frac{d\epsilon}{1+{\rm exp}(\frac{\epsilon}{T})},
\end{equation}
% {N_A k_B}
we calculated $M/B$.
As shown in Fig.~\ref{fig:kaiT}, the model predicts that $M/B$ also varies as $\sqrt{T}$ at low temperatures in $B=0.1$ T, whereas $M/B$ deviates downward in $B=5$ T.
The upturn in $M/B$ for $B//$c-axis is well reproduced in the present model, indicating that the upturn has the same origin as the NFL behavior in $C/T$. 
Furthermore, this agreement suggests that the Wilson ratio is nearly equal to one.
However, it is unclear why $M/B$ for $B//$a-axis at 2K is suppressed by $\sim 20$ \% compared to that for $B//$c-axis, although the $C/T$ shows no marked difference for either field direction.

% CeCu_{5.9}Au_{0.1}
The same analysis can be applied to another NFL system CeCu$_{5.9}$Au$_{0.1}$, where $C/T$ behaves as $C/T\propto-{\rm ln}(T/T_0)$ in zero field as reported by L\"{o}hneysen {\it et al}.~\cite{rf:Lo94CeCu6}
If we assume a rigid quasi-particle band expressed as $D_{\uparrow}(\epsilon)=D_{\downarrow}(\epsilon)=-u {\rm ln}(| \epsilon |/v)$, then this model yields $C/T=-(2\pi^2/3)u {\rm ln}(T/0.352 v)$ for $B=0$ and $C/T=-(2\pi^2/3)u {\rm ln}(\mu_{eff} B/k_B v)$ for $T=0$.
Comparing the equations with the corresponding experimental data,~\cite{rf:Lo94CeCu6} $u=0.089$ states$\cdot$J/K$^2$mol$\cdot$spin, $v=20.0$ K and $\mu_{eff}=1.92 \mu_B$ are obtained.
The calculated $C/T$ and $M/B$ (not shown) account for the following facts; (1) the broad maximum in $C/T$ vs $T$ for higher fields, which can be viewed as evidence of the peak in the quasi-particle-band DOS similarly as in CeNi$_2$Ge$_2$, and (2) the upturn in $M/B$ vs $T$ at low temperatures qualitatively.

%  AF fluctuation, SCR
We have demonstrated that the overall NFL behaviors of $C/T$ and $M/B$ for CeNi$_2$Ge$_2$ and CeCu$_{5.9}$Au$_{0.1}$ can be described as an anomalous sharp peak formed at the Fermi energy in the quasi-particle-band DOS within the Fermi liquid picture.
Since both systems are very close to the AF long-range ordering at $T=0$, the antiferromagnetic spin fluctuation (AFSF) is expected to play a role in forming the peak in the quasi-particle-band DOS.
In such a case, it is expected that the AF magnetic excitations are compressed to low energies,~\cite{rf:Gr96CPS} and the resulting abundance of low-lying excitations could lead to an enhancement of the quasi-particle mass or to a large $\gamma$ value.
Dealing with the AFSF based on the framework of the selfconsistent renormalized (SCR) theory, Moriya and Takimoto~\cite{rf:Mo95SCR} calculated several physical quantities, e.g., $C/T=a-b\sqrt{T}$ at low temperatures.
In the Ce$_{1-x}$La$_x$Ru$_2$Si$_2$ system, Kambe {\it et al.}~\cite{rf:Ka96SCR} reported that the experimental results can be understood to some extent in the frame of SCR theory.
Thus, the SCR calculation extended to the case in the applied fields is desirable for direct comparison with our results.

%  anisotropic hybridization
As an alternative explanation, anisotropic hybridization of 4{\it f}- and conduction electrons is possible.~\cite{rf:Ha87}
From the crystalline field (CF) analysis of the magnetic susceptibility above 50 K, the CF parameters of Ce$^{3+}$ ion $B_2^0=-12.2$ K, $B_4^0=0.05$ K and $B_4^4=-3.03$ K and the molecular-field coefficient $\lambda=-57$ mol/emu are obtained.
This large value of $\lambda$ indicates a strong Kondo effect and/or intersite AF correlation.
The latter effect appears as a broad $\chi _c$ maximum around 30 K, which cannot be explained in the frame of the CF model.
The CF parameters lead to the CF ground-state of $x|\pm5/2\rangle+\sqrt{(1-x^2)}|\mp3/2\rangle$ with $x\simeq0.90$ and the excited states lying above $\sim 200$ K.
Hybridization of this ground state with conduction electrons could lead to anomalous peaks in the density-of-states of the quasi-particle bands.~\cite{rf:Mi97,rf:Ik97}
For CeRu$_2$Si$_2$, one such peak is proposed as an origin of the observed metamagnetic anomaly at $B=7.7$ T.~\cite{rf:Ao97CRS}
Also for CeNi$_2$Ge$_2$, a peak located just at Fermi energy could be an origin of the observed peak.

\section*{Acknowledgments}
We thank Professor K. Miyake and M. D. Matsuda for the fruitful discussions. This work was supported by a Grant-in-Aid for Scientific Research from the Ministry of Education, Science, Sport and Culture.

\newpage
% ------------- Figures ------------
\begin{figure} %1
% \figureheight{0cm}
\caption{$\sqrt{T}$ dependence of $C/T$ measured in the field of $B//$c-axis and $B//$a-axis.\ \ \ \ \ \ \ \ \ \ \ \ \ \ \ \ \ \ \ \ \ \ \ \ \ \ \ \ }
\label{fig:CBTT}
\end{figure}

\begin{figure} %2
\caption{$B$-$T$ phase diagram.
The broad maximum in $C/T$-vs-$T$ curve ($T_{max}$) for both the experimental results ($\bullet$) and the model calculation (broken line through the origin) is shown.
In the shaded area, calculated $C/T$ becomes temperature-independent constant.
$T_l$ defined in ref. 4 is also plotted ($\circ$).}
\label{fig:BT}
\end{figure}

\begin{figure} %3
\caption{Magnetic field dependence of $C/T$ for $B//$c-axis and for $B//$a-axis.
The broken curve shows the model calculation predicted for $T=0$ (see text).}
\label{fig:CBTB}
\end{figure}

\begin{figure} %4
\caption{Calculated $C/T$ vs $\sqrt{T}$.
The inset shows a schematic of the energy dependence of the density-of-states for the model quasi-particle-bands in an applied field.
Fermi energy corresponds to $\epsilon = 0$.}
\label{fig:CBTTcalc}
\end{figure}

\begin{figure} %5
\caption{Calculated magnetic field dependence of $C/T$ in the model.~~~~~~~~~~~~~~~~~~~~~~~~~~~~~~~~~~~~~~~~~~~~~~~~~~~~~~~~}
\label{fig:CBTBcalc}
\end{figure}

\begin{figure} %6
\caption{Magnetic field dependence of the entropy $S$ determined from the magnetocaloric effect measurement and integration of $C/T$-vs-$T$ curve.
The model calculation is shown as a broken curve.}
\label{fig:SB}
\end{figure}

\begin{figure} %7
\caption{$\sqrt{T}$ dependence of the magnetic susceptibility $\chi\equiv M/B$ for $B$//c-axis and $B$//a-axis.
The full and broken curves are for the model calculations.}
\label{fig:kaiT}
\end{figure}

\end{document}